# Implications of beam phase and RFSUM measured near transition

Xi Yang and James MacLachlan

*Fermi National Accelerator Laboratory*

Box 500, Batavia IL 60510

**Abstract**

Understanding the transition-crossing process is crucial for improving Booster performance at high intensity. The synchronous phase appears to drop toward 90° right after transition regardless of beam intensity, more so at higher beam intensity. The implication is that the effective rf voltage (RFSUM) will run into a limit right after transition when the synchronous phase reaches 90° for high intensity beam. A reduction in RFSUM is also observed at the same time. Solutions, such as raising the rf voltage during the transition period or controlling the RFSUM reduction by increasing longitudinal emittance before transition, are potentially important for high intensity operation.

## Introduction

It is important to have diagnostic tools to measure the transition time and monitor the details of the synchronous phase and effective rf voltage changes during the transition-crossing rf manipulations. By correlating experimental observations with Booster performance, Booster running conditions can be optimized, especially for high intensity beam. A synchrotron phase detector in the low level rf room of the Booster measures the phase between the rf fundamental component of the beam current and the effective rf voltage with a time resolution of one microsecond.[1] The RFSUM signal gives the effective rf voltage acting on the beam.[2] The combination of the synchrotron phase detector with the RFSUM signal provides not only the phase motion of the beam relative to the rf waveform but also the energy loss per turn during the transition crossing period.

## Observations

The synchronous phase was measured at extracted beam intensities of $2.6\times10^{12}$ protons, $3.3\times10^{12}$ protons, and $5.1\times10^{12}$ protons using the synchrotron phase detector. The results are shown in Fig. 1. The duration of the transition phase jump, about 16 μs, appears to be independent of beam intensity. From the data shown in Fig. 1, the synchronous phase at these extracted beam intensities is plotted in Fig. 2(a) at three different times, right before transition, right after transition, and when the synchronous phase has reached its minimum at about 60 μs to 70 μs after transition. The phase jump, the difference between the red curve and the black curve in Fig. 2(a), and the minimum phase change, the difference between the green curve and the black curve, are plotted in Fig. 2(b). The RFSUM signal, taken at the extracted beam intensity of $4.1\times10^{12}$ protons, is shown in Fig. 3. The transition crossing period, indicated by red arrows, is about 14 μs. This result is consistent with the one from the synchronous phase measurement. An important feature appearing in both the synchronous phase measurement and the RFSUM signal is that the synchronous phase drops closer to 90° right after transition (see Fig. 1) at the time of the RFSUM reduction, indicated by the green arrows in Fig. 3. Also, the energy loss per Booster turn can be obtained from the combination of the synchrotron phase detector signal and the RFSUM signal.[3] The energy loss per turn during the period near transition is shown in Fig. 4 for three extracted beam intensities of $2.6\times10^{12}$ protons, $3.3\times10^{12}$ protons, and $5.1\times10^{12}$ protons.

**Comments**

The duration of the transition phase jump is estimated to be about 14 μs to 16 μs from both the synchronous phase measurement and the RFSUM signal, consistent with an estimate based upon the Q of Booster rf cavities, about 800 at transition.[4] It takes about 9.5 Booster turns for the phase jump, about 15.4 μs since the revolution period at transition is 1.615 μs. The phase jump is dependent upon the beam intensity, and is shown by the black curve and green curve in Fig. 2(a). The synchronous phase is closer to 90° when the beam intensity goes higher, both before and after transition. Note, however, the red curve in Fig. 2(a) indicates that the higher the beam intensity is, *the closer the synchronous phase is to 180° right after transition*. These two observations seem consistent with the observation that the phase jump *at transition* is *independent* of the beam intensity, as shown in the red curve of Fig. 2(b). However, the difference between the synchronous phases, which are closest to 90° before and some tens of microseconds after transition, is *dependent* upon the beam intensity, as shown in the black curve of Fig. 2(b). Thus, the higher the beam intensity, the more effective rf voltage is required. The effective rf voltage (RFSUM) at transition could become the limiting factor for the highest beam intensity in Booster.

The RFSUM reduction after transition, shown by the green arrows in Fig. 3, is certainly related to the observation of the synchronous phase dropping toward 90° at a similar time, consistent with the increased energy loss per turn during the transition period, as shown in Fig. 4. This conclusion implies that it should help to get higher intensity beam after transition if the RFSUM dip at transition is removed or reduced, perhaps by increasing the effective rf voltage during the transition period or by intentionally increasing the longitudinal emittance to reduce the energy loss caused by bunch shortening near transition.

References:

[1] X. Yang and R. D. Padilla, A Synchronous Phase Detector for the Fermilab Booster, FERMILAB-TM-2234.

[2] J. Steimel, Fermilab / Accelerator Division / RF Department / HLRF (FanBack / FanOut) Document. (1992)

[3] X. Yang, C. Ankenbrandt, and J. Norem, Experimental Estimate of Beam Loading and Minimum rf Voltage for Acceleration of High Intensity Beam in the Fermilab Booster, FERMILAB-TM-2237, submitted

[4]K. C. Harkay, A Study of Longitudinal Instabilities and Emittance Growth in the Fermilab Booster Synchrotron, PhD thesis. (1993)

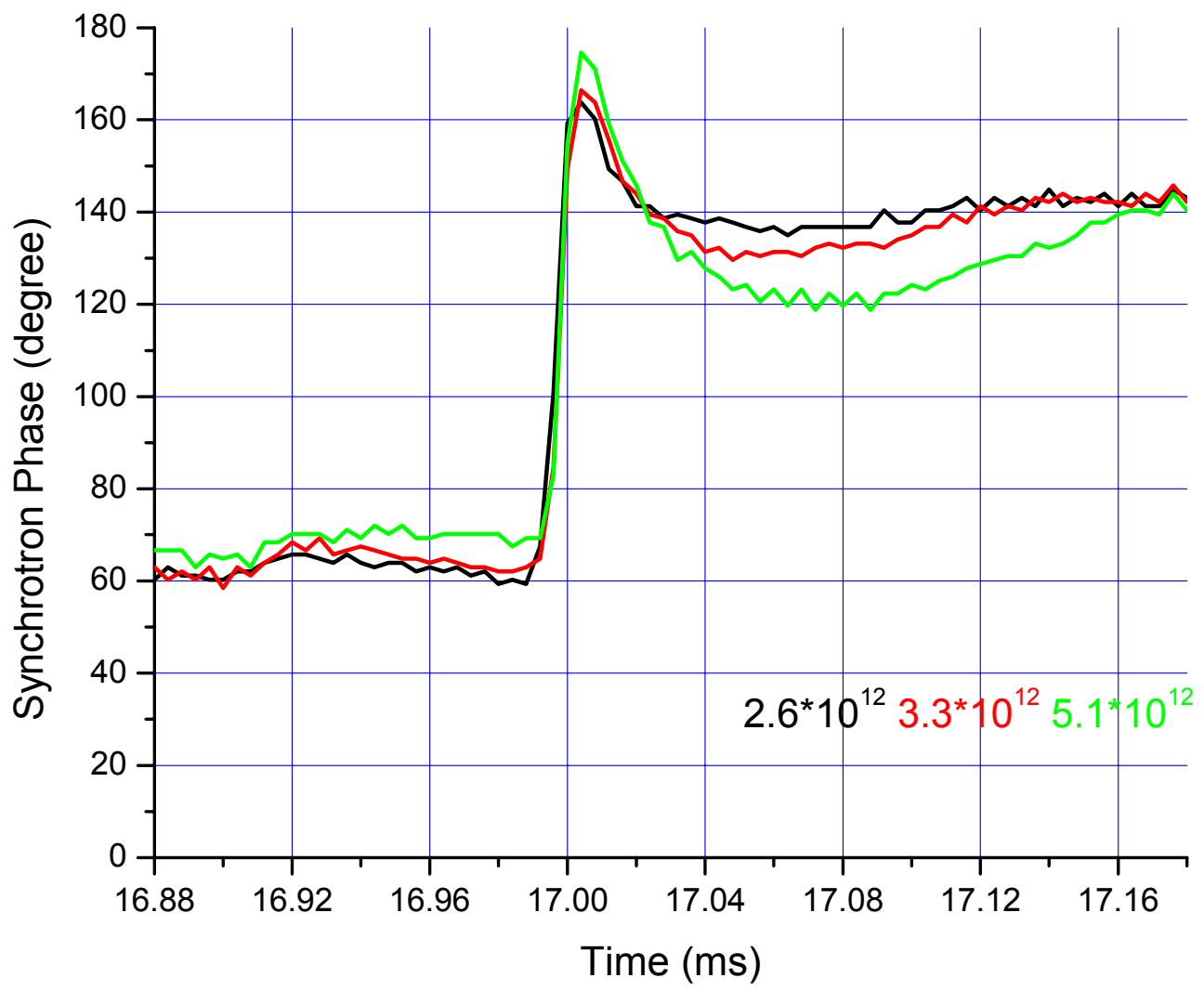

Fig. 1

Fig. 1. The synchronous phase measured at three extracted beam intensities of $2.6\times10^{12}$ protons (the black curve), $3.3\times10^{12}$ protons (the red curve), and $5.1\times10^{12}$ protons (the green curve).

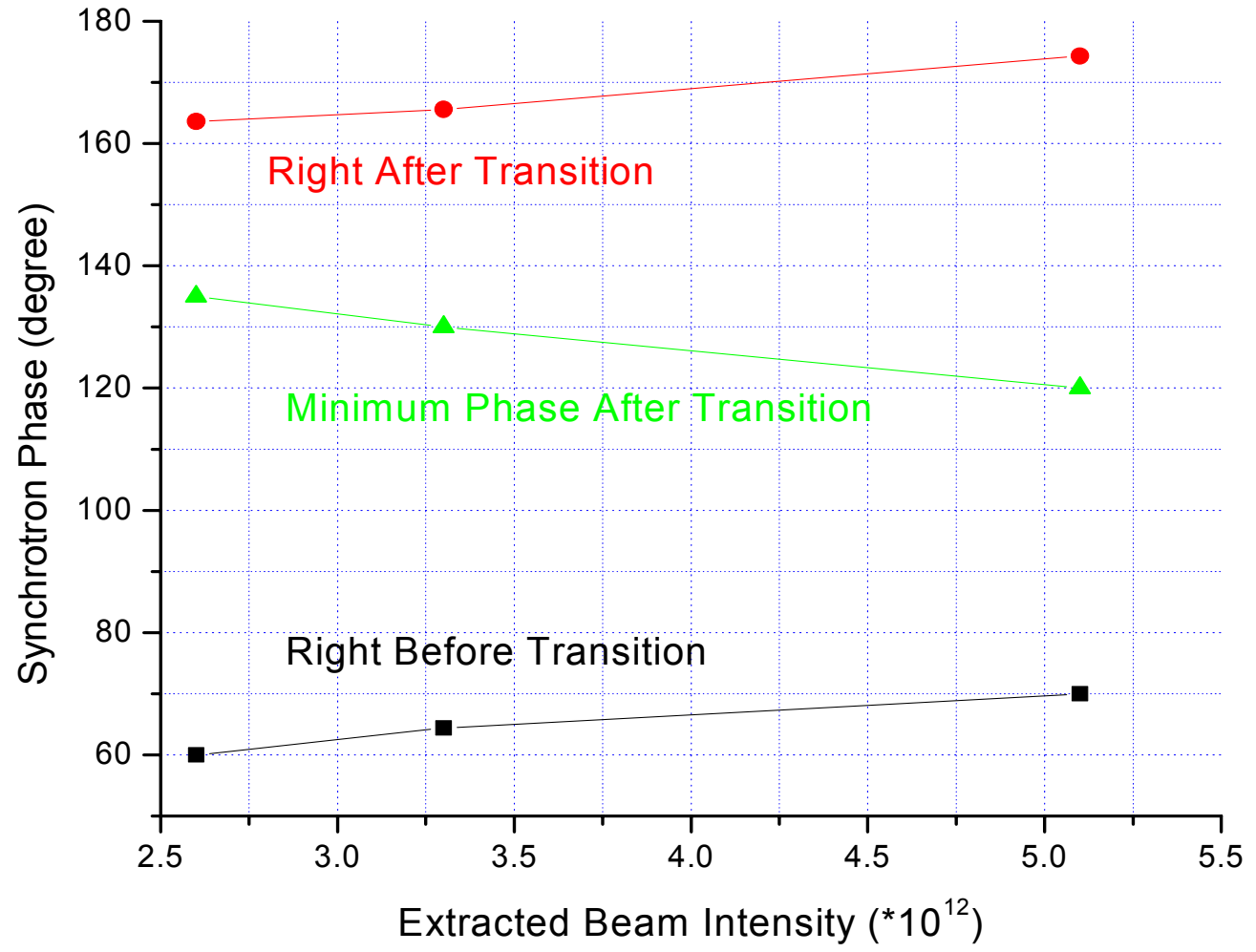

Fig. 2(a)

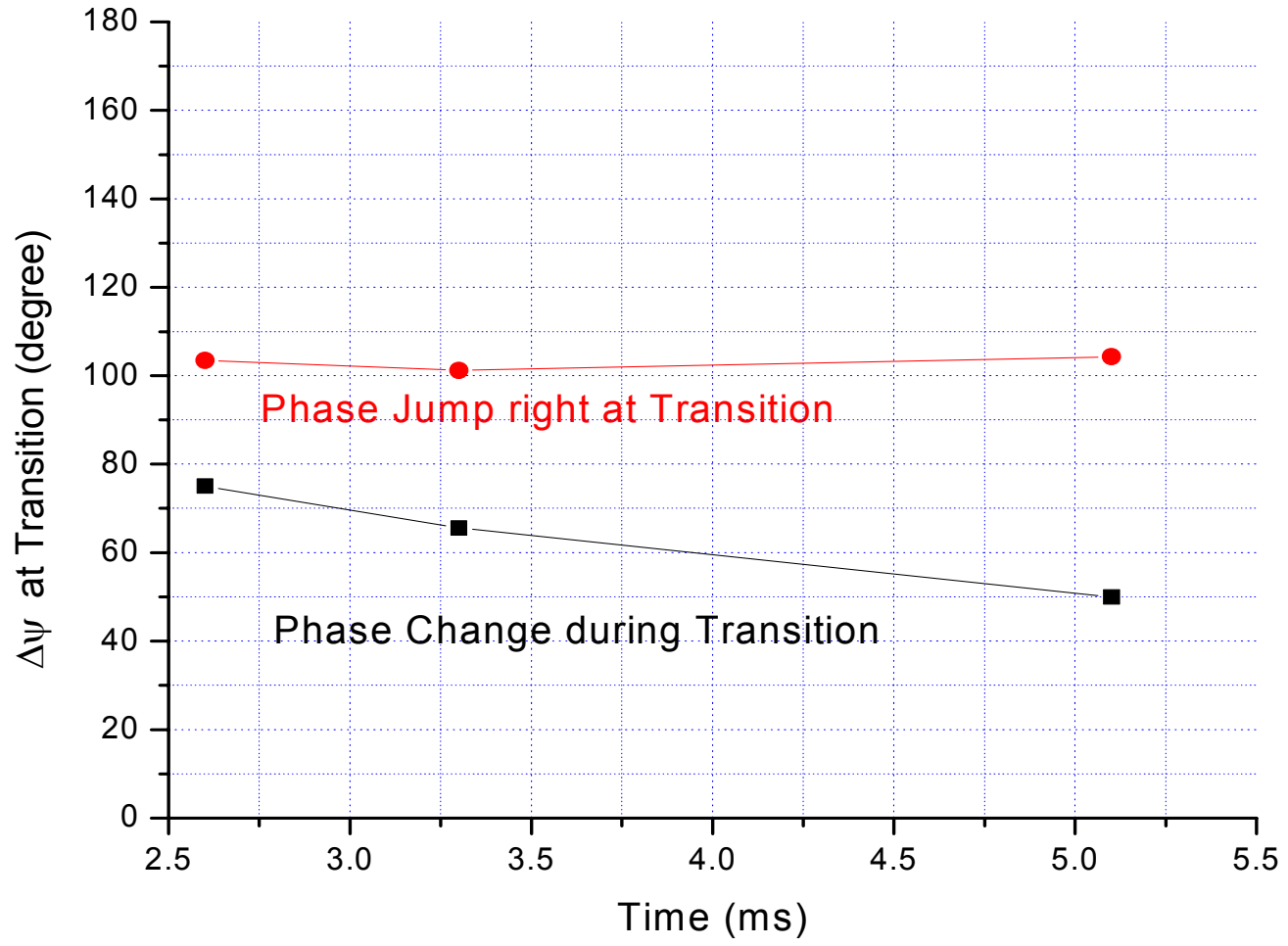

Fig. 2(b)

Fig. 2(a) the black curve represents the synchronous phase right before transition, the red curve represents the synchronous phase right after transition, and the green curve represents the synchronous phase at about 60 μs -70 μs after transition when it reached its minimum.

Fig. 2(b) the red curve represents the synchronous phase jump right across transition (the difference between the red curve and black curve in Fig. 2(a)), the black curve represents the synchronous phase change during transition (the difference between the green and black curve in Fig. 2(a)).

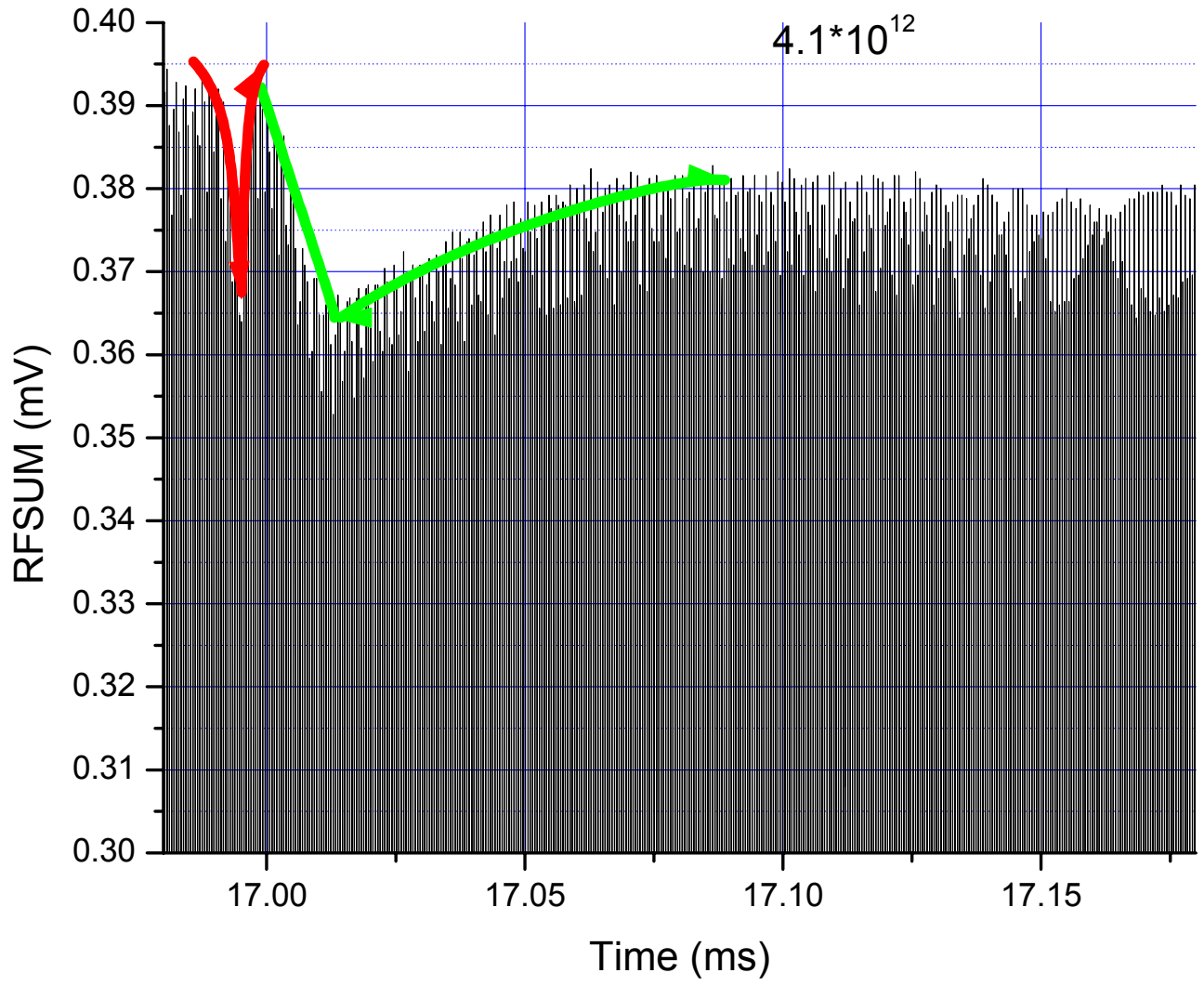

Fig. 3

Fig. 3 the RFSUM signal during the transition period. The RFSUM reduction at transition is indicated by red curves. The RFSUM reduction after transition is indicated by the green curves.

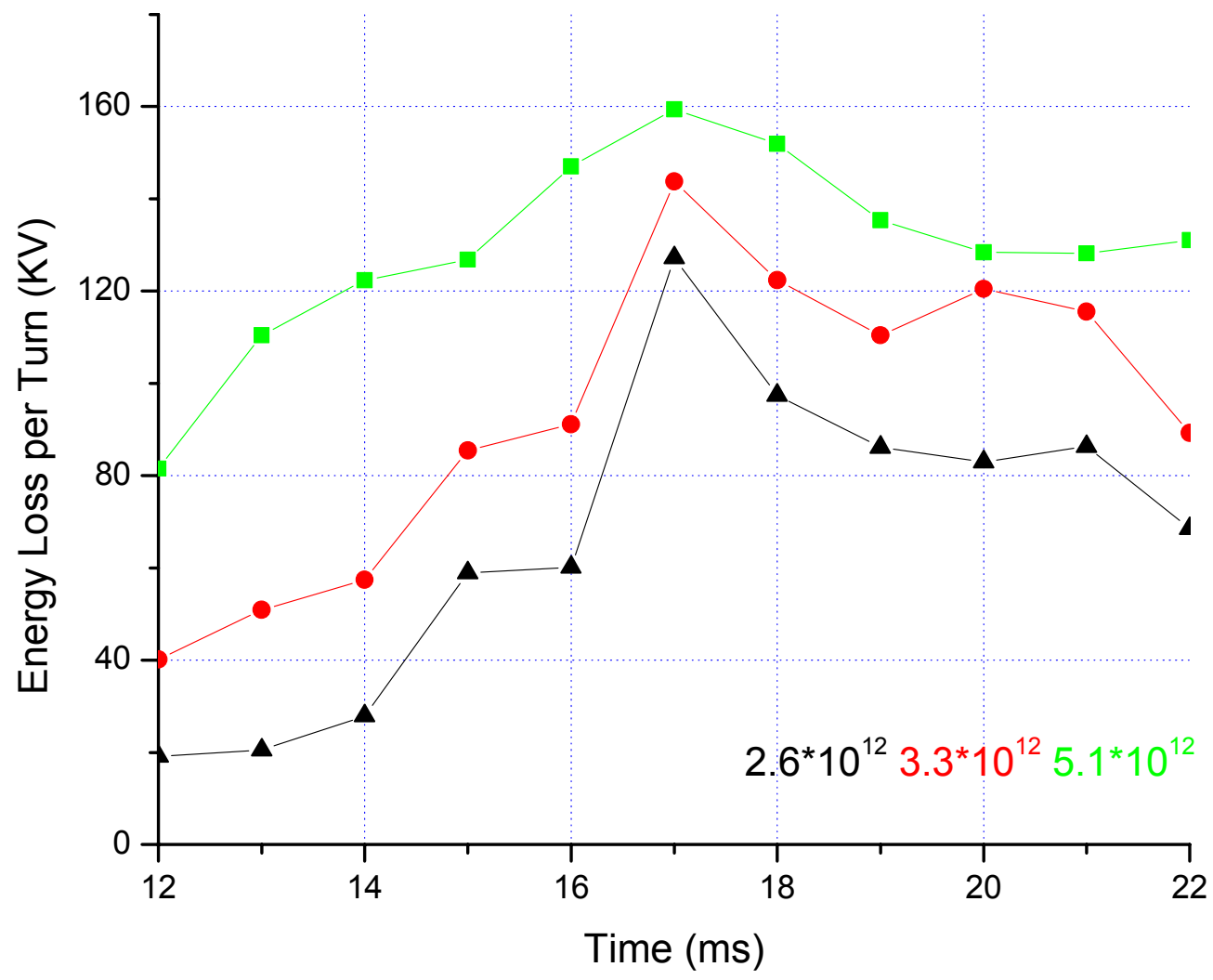

Fig. 4

Fig. 4. the estimate of the beam energy loss per turn in a Booster cycle near transition at three extracted beam intensities of $2.6\times10^{12}$ protons (the black curve), $3.3\times10^{12}$ protons (the red curve), and $5.1\times10^{12}$ protons (the green curve).